\documentclass[footinbib,aps,pra,reprint,superscriptaddress,nopacs]{revtex4-1}

\usepackage{amsmath,amssymb,braket,upgreek,bm,verbatim}
\usepackage[absolute]{textpos}
\usepackage{physics, graphicx, tabularx,braket}
\usepackage{xcolor}

\usepackage[absolute]{textpos}
\usepackage{multirow}

\definecolor{darkblue}{rgb}{0,0,0.5}

\usepackage{hyperref}
\hypersetup{
colorlinks=true,
linkcolor=black,
filecolor=blue,
citecolor=darkblue,  
urlcolor=black,
}

\begin{document}
\title{CMOS photonic integrated source of ultrabroadband polarization-entangled photons}

\author{Alexander Miloshevsky}
\email{miloshevskya@ornl.gov.}
\affiliation{Quantum Information Science Section, Computational Sciences and Engineering Division, Oak Ridge National Laboratory, Oak Ridge, Tennessee 37831, USA}

\author{Lucas M. Cohen}
\thanks{These authors contributed equally to this work.}
\author{Karthik V. Myilswamy}
\thanks{These authors contributed equally to this work.}
\affiliation{School of Electrical and Computer Engineering and Purdue Quantum Science and Engineering Institute, Purdue University, West Lafayette, Indiana 47907, USA}

\author{Muneer Alshowkan}
\affiliation{Quantum Information Science Section, Computational Sciences and Engineering Division, Oak Ridge National Laboratory, Oak Ridge, Tennessee 37831, USA}

\author{Saleha Fatema}
\affiliation{School of Electrical and Computer Engineering and Purdue Quantum Science and Engineering Institute, Purdue University, West Lafayette, Indiana 47907, USA}

\author{Hsuan-Hao Lu}
\affiliation{Quantum Information Science Section, Computational Sciences and Engineering Division, Oak Ridge National Laboratory, Oak Ridge, Tennessee 37831, USA}

\author{Andrew M. Weiner}
\affiliation{School of Electrical and Computer Engineering and Purdue Quantum Science and Engineering Institute, Purdue University, West Lafayette, Indiana 47907, USA}

\author{Joseph M. Lukens}
\affiliation{Quantum Information Science Section, Computational Sciences and Engineering Division, Oak Ridge National Laboratory, Oak Ridge, Tennessee 37831, USA}
\affiliation{Research Technology Office and Quantum Collaborative, Arizona State University, Tempe, Arizona 85287, USA}

\date{\today}

\begin{textblock}{13.3}(1.4,15)
\noindent\fontsize{7}{7}\selectfont \textcolor{black!30}{This manuscript has been co-authored by UT-Battelle, LLC, under contract DE-AC05-00OR22725 with the US Department of Energy (DOE). The US government retains and the publisher, by accepting the article for publication, acknowledges that the US government retains a nonexclusive, paid-up, irrevocable, worldwide license to publish or reproduce the published form of this manuscript, or allow others to do so, for US government purposes. DOE will provide public access to these results of federally sponsored research in accordance with the DOE Public Access Plan (http://energy.gov/downloads/doe-public-access-plan).}
\end{textblock}

\begin{abstract}
We showcase a fully on-chip CMOS-fabricated silicon photonic integrated circuit employing a bidirectionally pumped microring and polarization splitter-rotators tailored for the generation of ultrabroadband ($>$9~THz), high-fidelity (90--98\%) polarization-entangled photons. Spanning the optical C+L-band and producing over 116 frequency-bin pairs on a 38.4~GHz-spaced grid, this source is ideal for flex-grid wavelength-multiplexed entanglement distribution in multiuser networks.
\end{abstract}

\maketitle

Quantum networks of the future~\cite{Wehner2018} must be flexible to distribute  entanglement on demand to multiple end-users, adapt to user resource requirements, and maneuver unexpected disruptions to communication channels. To this end, recent experiments combining broadband polarization entanglement with wavelength-selective switches (WSSs) chart a promising path forward~\cite{Lingaraju2021, Appas2021, Alshowkan2021}. Leveraging concepts proven in classical ``flex-grid'' or ``elastic'' optical networking~\cite{Gerstel2012}, the center wavelength, bandwidth, and lightpath of any quantum demand can be reconfigured adaptively. In this paradigm, broadband and compact polarization-entangled sources---ideally via scalable, CMOS integrated photonics---are of critical importance. 
Multiple approaches for the on-chip generation of polarization-entangled photons have been proposed and demonstrated; examples employing spontaneous four-wave mixing (SFWM) in silicon include polarization rotation in a single waveguide~\cite{Matsuda2012}, parallel waveguides combined with a two-dimensional grating coupler~\cite{Olislager2013}, and multi-spatial-mode interactions~\cite{Feng2019b}, while type-II parametric downconversion in AlGaAs has also proven extremely successful~\cite{Baboux2023}.

Ubiquitous in silicon photonics, microring resonators (MRRs) enable resonantly enhanced SFWM for pair generation with higher brightness (flux per unit bandwidth) than single-pass waveguide structures, all with a more compact footprint~\cite{Moody2020}. While producing frequency-bin entanglement automatically~\cite{Kues2019, Lu2023b}, the direct generation of polarization entanglement in MRRs is complicated by the distinct spatial profiles and effective indices for orthogonally polarized modes, leading to mismatched spectral resonances for transverse-electric (TE) and transverse-magnetic fields (TM). 
This situation has motivated designs in which an MRR is placed in a fiber Sagnac loop to convert copolarized but counterpropagating amplitudes into a polarization-entangled output~\cite{Suo2015, Wen2023}---a solution that unfortunately sacrifices compactness for functionality.
Accordingly, integrated sources of polarization entanglement have so far been limited either by 
(i)~adopting a nonresonant on-chip solution with lower efficiency or (ii)~enlisting efficient MRRs but supplementing them with off-chip fiber-optic manipulation.

\begin{figure}[t]
    \centering
    \includegraphics[width=\linewidth]{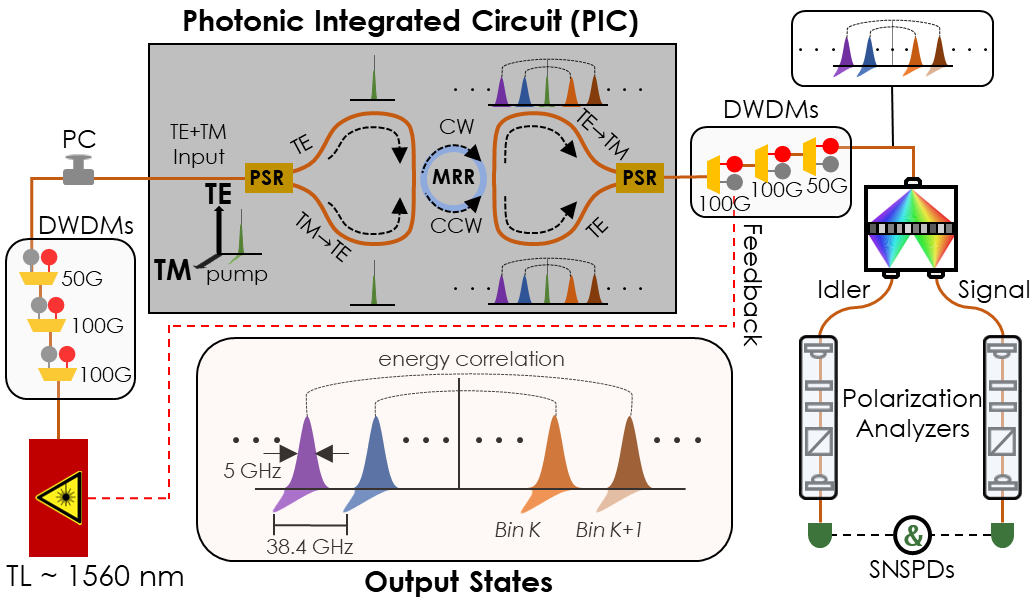}
    \caption{Experimental setup including conceptual diagrams of the polarization-frequency states at various stages. TE: transverse electric; TM: transverse magnetic; MRR: microring resonator; TL: continuous-wave tunable laser; DWDM: dense wavelength-division multiplexer; PC: polarization controller; PSR: polarization splitter-rotator; SNSPD: superconducting nanowire single-photon detector.}
    \label{fig:Drawings}
\end{figure}

Here we propose a novel solution combining bidirectional pumping with integrated polarization splitter-rotators for a complete on-chip, MRR-based, CMOS-fabricated polarization-entangled source. We verify high-fidelity operation through quantum state tomography (QST) of 116 individual channel pairs and various channel groupings across the full optical C+L-band, confirming the viability of flexible bandwidth allocations as well as the tradeoff between fidelity and flux due to multipair emission and polarization-mode dispersion (PMD). 

Figure~\ref{fig:Drawings} illustrates the experimental setup. 
A continuous-wave laser (Santec TSL-570) pumps the chip, tuned to the ring resonance at $\sim$1559.85~nm. To attenuate background noise in the pump laser, we incorporate three dense wavelength division multiplexers (DWDMs): two with a 100~GHz passband and one with a 50~GHz passband, all centered around 1559.81~nm. Following the DWDMs, a polarization controller (PC) manages the ratio of TE and TM polarization coupled into the chip ($\sim$10~dBm estimated on-chip power).

The die, designed for and manufactured by the AIM Photonics multi-project wafer service~\cite{Fahrenkopf2019}, leverages 
polarization-independent spot-size converters to couple light of both polarizations onto the chip. A polarization splitter-rotator (PSR) spatially separates TE and TM inputs, rotating the TM portion to TE. These pump photons bidirectionally couple into a racetrack-shaped MRR with $500$~nm-wide single-mode waveguides throughout and adiabatic curves and directional couplers in the coupling sections. 
The MRR is designed with a free spectral range of 38.4~GHz and intrinsic (loaded) $Q$ factors of $3.7\times 10^5$ ($3.7\times 10^4$);  experimentally, loaded $Q$ factors are 
$\sim$$3.4\times 10^4$ near the pump resonance. 
Resistive microheaters are embedded within the MRR and in each output path for frequency and phase tuning, 
but are not employed here.

After extracting the TE-polarized frequency-bin-entangled photons generated via SFWM in both directions of the MRR, a second PSR reverts one pathway to TM polarization before recombining with the unrotated TE pathway. Ideally, the output is an equal coherent superposition of two counterpropagating SFWM processes, where signal and idler photons form a biphoton frequency comb with a frequency spacing (38.4 GHz) and linewidth ($\sim$5~GHz full-width at half-maximum) defined by the MRR. Each pair of energy-matched resonances manifests a polarization Bell state $\ket{\Phi^+}= \frac{1}{\sqrt{2}} (\ket{H_{\omega_k}H_{\omega_{-k}}} + \ket{V_{\omega_k}V_{\omega_{-k}}})$, where TE (TM) is identified with horizontal $H$ (vertical $V$) polarization and the integer $k\geq 1$ denotes the $k^{\mathrm{th}}$ frequency-bin pair (number of resonances away from the pump). 

Upon exiting the chip, another set of three DWDMs, replicating the prior configuration, isolates the newly generated broadband entangled photons from the residual pump light, with an estimated pump suppression of $\sim$68~dB. Additionally, we implement a feedback loop that monitors the optical power in the pass channel of the first output DWDM, ensuring the pump aligns with the intended cavity resonance in the presence of real-time thermal drift. Finally, we employ a Fourier-transform pulse shaper (Finisar Waveshaper 4000B) as a WSS to route two frequency bins, one from the signal side of the spectrum and one from the idler, to two separate optical fibers for subsequent state characterization. Two-qubit polarization QST is carried out using a pair of motorized polarization analyzers (consisting of a quarter-wave plate, a half-wave plate, and a polarizing beamsplitter cube) followed by photon detection with superconducting nanowire single-photon detectors (Quantum Opus) connected to a photon-counting module (PicoQuant TimeHarp).

\begin{figure*}[ht]
    \centering
    \includegraphics[width=\textwidth]{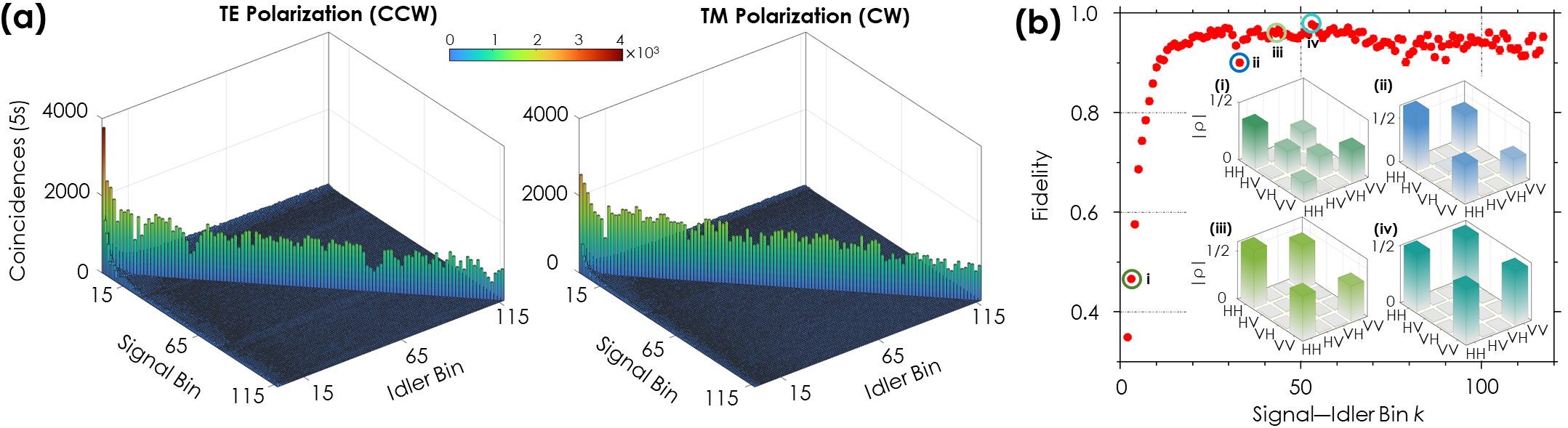}
    \caption{(a) Unidirectionally pumped JSIs measured for bins $k\in\{2,...,117\}$ (1~ns coincidence window). (b) Bayesian state fidelities for the 116 energy-matched frequency bins. Density matrix absolute values for select bins (i-iv).}
    \label{fig:Results}
\end{figure*}

We first characterize the MRR by measuring the joint-spectral intensity (JSI) under unidirectional pumping. We tune the PC to excite only the TE or TM polarization, bypass both polarization analyzers, and program the pulse shaper to scan through a total of $116\times116$ frequency-bin combinations for coincidence measurements ($k\in\{2,...,117\}$) [Fig.~\ref{fig:Results}(a)]; the $k=1$ bin pair is clipped by the DWDMs. 
The $\sim$10 closest pairs experience higher accidental coincidences from residual pump light, which would necessitate additional DWDMs for further noise suppression. 
As examples, the measured coincidence-to-accidental ratios (CARs) for TE (TM) inputs increase from 1.3 (1.8) at $k=1$ to 33 (43) at $k=11$, and finally reach 260 (360) at $k=117$.
From heuristic fitting based on coincidence and single-detector count rates~\cite{Alnas2022}, we estimate on-chip pair generation rates of $\sim$10$^7$~s$^{-1}$ per bin. No significant decrease in coincidence counts---relative to the fluctuations across the spectrum---was observed within the 9~THz examined, consistent with theoretical modeling~\cite{Chembo2016} predicting a 3~dB SFWM bandwidth exceeding 16~THz. 


We conduct polarization QST on these 116 energy-matched bin pairs after adjusting the pump polarization to balance output amplitudes from the two counterpropagating processes. We scan through 36 polarization projections, collecting coincidence counts (30~s per projection) for subsequent Bayesian state analysis~\cite{BlumeKohout2010, Lukens2020, Lu2022b}. For each bin pair, we infer 1024 density matrix samples from the gathered dataset and apply local rotations numerically to align the measured reference frame with that of the chip~\cite{Alshowkan2022}. Fidelities with respect to $\ket{\Phi^+}$ are depicted in Fig.~\ref{fig:Results}(b). As expected, fidelities are lower for channels closer to the pump frequency [cf. density matrix (i)] 
but steadily improve with increasing channel separation, reaching a plateau within the range of approximately 90--98\% across the rest of the spectrum. Three example density matrices (ii)-(iv) are shown in Fig.~\ref{fig:Results}(b) for bins at the minimum, average, and maximum fidelity, respectively, over the range $k\in[11,117]$.  
The cause of the fluctuations in this region is attributed to polarization or coupling drifts during the full measurement procedure which could be rectified with a polarization feedback loop or affixing the fiber array to the die. The average fidelity for all bins is 92(9)\%, increasing to 94(2)\% when excluding $k\in[2,10]$. These results mark a record number of fully characterized polarization-entangled channels from an integrated source, paving the way for ultracompact CMOS sources for quantum networking. Further measurements in the joint polarization-frequency domain~\cite{Lu2023a} can help confirm that our design also provides true hyperentanglement in both degrees of freedom, valuable for quantum communication protocols such as dense coding~\cite{Barreiro2008} and entanglement distillation~\cite{Simon2002, Sheng2010}.

Given the combination of narrow bin spacing and full C+L-band coverage, this source is well suited for dense multiplexing in the flex-grid paradigm of entanglement distribution, in which WSSs dynamically partition a broadband biphoton spectrum into numerous slots of energy-correlated bands and subsequently route them to multiple destination nodes (users) to establish entanglement links~\cite{Lingaraju2021, Appas2021, Alshowkan2021}. 

In the simplest scenario, assigning one frequency bin pair to each pair of users could support high-fidelity entanglement for more than 100 pairs of nodes [cf. Fig.~\ref{fig:Results}(b)], constrained only by the WSS passband and number of output fibers. As certain users contend with increased link losses or seek additional services, we have the flexibility to group adjacent bins together for reallocation, effectively enhancing the coincidence rate. However, such bandwidth expansion can easily decrease fidelity: unless otherwise limited by background noise, increasing flux reduces CAR due to higher multipair-induced accidentals~\cite{Alnas2022}, and wider-bandwidth channels are more sensitive to any PMD present.

\begin{figure*}[tb!]
    \centering
    \includegraphics[width=\textwidth]{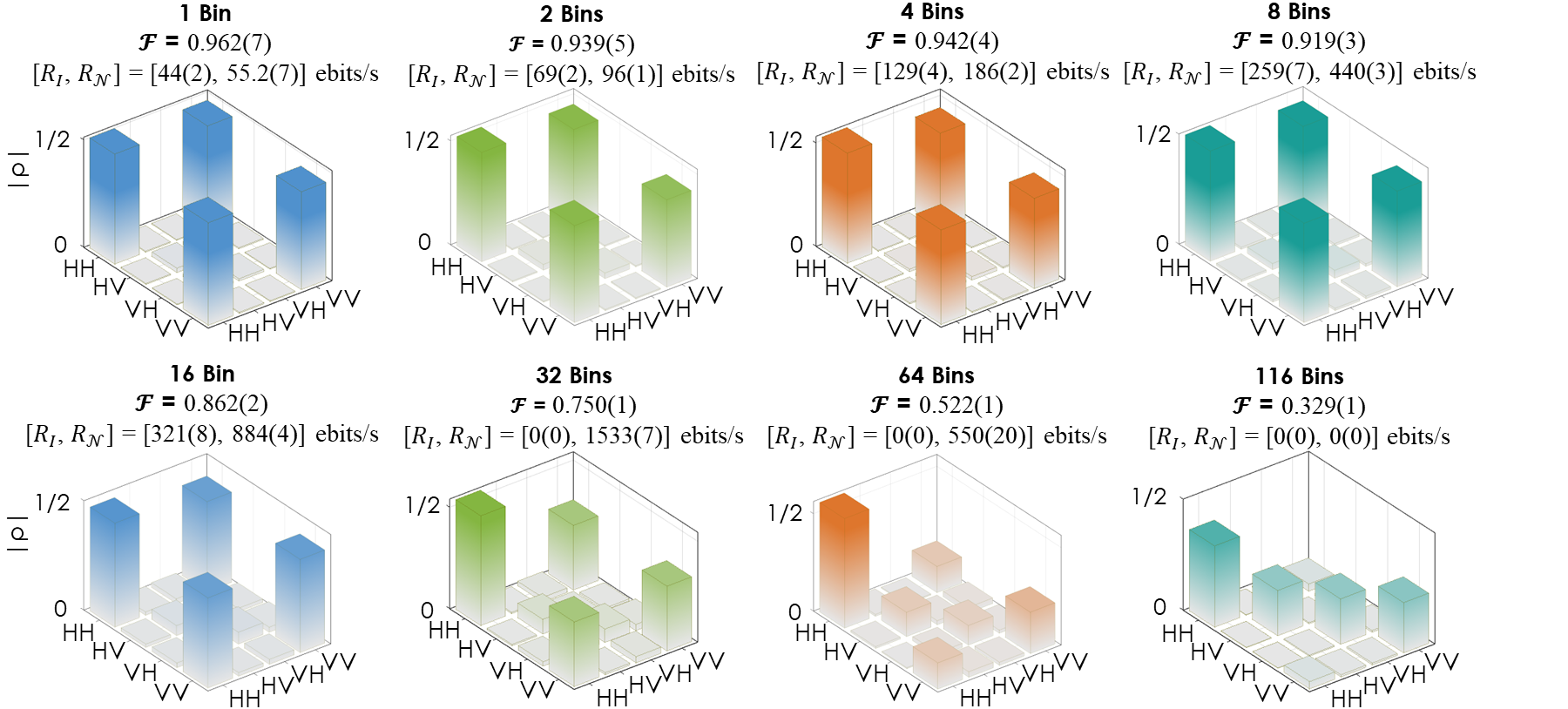}
    \caption{Experimentally measured density matrix absolute values for $\{1, 2, 4, 8, 16, 32, 64, 116\}$ grouped frequency bins, their fidelity, and the lower and upper bound of the distillable entanglement, all obtained from Bayesian QST.}
    \label{fig:DM}
\end{figure*}

Consequently, achieving a delicate balance between optimizing throughput and maintaining entanglement quality above a predefined, application-specific threshold becomes a crucial area for investigation. Figure~\ref{fig:DM} offers an initial glimpse into this exploration, illustrating how bin grouping impacts state quality as evaluated through polarization QST. Our approach involves initially passing one bin on each side of the spectrum ($k = 58$) and symmetrically widening the filter passband, resulting in output states composed of $\{2, 4, 8, 16, ...\}$ contiguous frequency bins. As anticipated, fidelity decreases with larger bin groupings, while the coincidence rate rises---a tradeoff which we can quantify through the entangled bit rate (EBR) defined as the distillable entanglement times the coincidence rate~\cite{Alshowkan2022}. Lower and upper bounds on EBR (ebits/s), $[R_I, R_\mathcal{N}]$, can be obtained from the coherent information~\cite{Schumacher1996} and log-negativity~\cite{Vidal2002}, respectively, as included in Fig.~\ref{fig:DM}. In general, EBR increases as bins are grouped together due to the rising flux but eventually decreases as the polarization state becomes increasingly mixed. 

For insight into the nonidealities present in our system,
we propose a model 
of a two-photon polarization state that incorporates accidental coincidences and imbalance and decoherence between $\ket{HH}$ and $\ket{VV}$ amplitudes. This theoretical density matrix is 
$\rho_{T} = \lambda\rho_{AB}+(1-\lambda)\rho_{A}\otimes\rho_{B}$,
where $\rho_{AB}$ is the density matrix of two time-correlated photons subject to decoherence and $H/V$ imbalance, given by
\begin{equation}
    \begin{split}
    \rho_{AB} = &h\ket{HH}\bra{HH}+g\sqrt{h(1-h)}\ket{HH}\bra{VV} \\
               &+g\sqrt{h(1-h)}\ket{VV}\bra{HH}+(1-h)\ket{VV}\bra{VV}.
    \end{split}
\end{equation}
Here $\lambda\in[0,1]$ parametrizes multipair emission (ideally 1), $h \in$ [0,1] denotes the relative probability of $\ket{HH}$ compared to $\ket{VV}$ (ideally 0.5), and $g\in [0,1]$ quantifies the quantum coherence (ideally 1)---without loss of generality assumed real and positive because of the local rotations performed to align the empirical density matrix with $\ket{\Phi^+}$~\cite{Alshowkan2022}. 
Given the starting bin at $k=58$ (far from the pump) we assume accidentals are dominated by uncorrelated pairs rather than background; the accidentals contribute to the state in the form of a product of marginal density matrices $\rho_A = \Tr_B \rho_{AB} = h\ket{H}\bra{H} + (1-h)\ket{V}\bra{V} = \Tr_A \rho_{AB} = \rho_B$.
The overlap of the model state with the ideal $\ket{\Phi^+}$ is 
$F_T = \braket{\Phi^+|\rho_T|\Phi^+} = \frac{1}{2} - h(1-\lambda)(1-h) + g\lambda\sqrt{h(1-h)}$

While simple, this model matches experimental observations well; numerical fits of
$\rho_T(\lambda,h,g)$ to the experimentally determined density matrices in Fig.~\ref{fig:DM} agree with greater than 95\% fidelity in all cases. 
Although the extracted $H/V$ imbalance varies slightly ($h\in[0.52, 0.68]$), 
its impact on fidelity is negligible compared to that of accidentals; $\lambda=0.95$ at 1 bin but by 32 bins has dropped sufficiently ($\lambda=0.71$) to reduce $R_I$ to zero---incidentally, $R_\mathcal{N}$ is still large, highlighting the comparatively wide spread in these lower and upper bounds. Overall, this scaling emphasizes and quantifies the inherent fidelity-flux tradeoff for these sources~\cite{Alshowkan2021, Alnas2022}, from which a desired operating point can be chosen for a specific application.

On the coherence front, $g>0.92$ until the 64- and 116-bin cases, where $g=0.59$ and $g=0.39$, respectively. With polarization-dependent delay in the WSS rated at $<$0.5~ps (and measured at $\sim$0.1~ps with a PMD tester), it is unsurprising to see such a reduction in coherence for broad bandwidths of several THz; quantitative predictions would require further tests to determine the orientation of the PMD principal axes with respect to the state's $H/V$ modes~\cite{Antonelli2011}.
However, it is important to note that although sensitive to off-chip PMD, our design is automatically robust to small path length mismatch on the chip. Any difference in length for the arms leaving and entering the PSRs imparts a delay between the output $H$ and $V$ contributions equal for both photons, which cancels out on the state $\ket{HH}+\ket{VV}$ (as long as the mismatch is much shorter than the pump laser coherence time~\cite{Shtaif2011}). 
Such PMD cancellation is a fascinating feature of entangled photon pairs~\cite{Antonelli2011} and an additional source of robustness in our source design; despite not being a ``Sagnac'' in the sense of truly common paths, insensitivity to small path variations is still obtained.
In summary, we propose and test a novel, compact, CMOS-fabricated photonic integrated circuit for the generation of high-fidelity polarization-entangled photons over an ultrabroad band. Our source has potential application for quantum networking as an entangled source and can tune the EBR through flex-grid networking.

\begin{acknowledgments}
Preliminary results were presented at IEEE IPC 2023 as paper number PD5. This work was performed in part at Oak Ridge National Laboratory, operated by UT-Battelle for the U.S. Department of energy under contract no. DE-AC05-00OR22725. Funding was provided by the U.S. Department of Energy, Office of Science, Advanced Scientific Computing Research (Early Career Research Program, ReACT-QISE); National Science Foundation (2034019-ECCS, 1747426-DMR); Air Force Research Laboratory (FA8750-20-P-1705).
\end{acknowledgments}

\end{document}